\documentclass[12pt]{article}
\usepackage{epsfig,here}
\usepackage{graphicx}
\begin{document}
\renewcommand{\theequation}{\arabic{section}.\arabic{equation}}

\newcommand{\bea}{\begin{eqnarray}}
\newcommand{\eea}{\end{eqnarray}}
\newcommand{\fs}{\;  .}
\newcommand{\co}{\;  ,}
\newcommand{\nn}{\nonumber \\}
\newcommand{\scs}{\co \;}
\newcommand{\sem}{ \; \ ; \;\;}
\newcommand{\la}{\langle}
\newcommand{\ra}{\rangle}
\newcommand{\GeV}{\mbox{GeV}}
\newcommand{\MeV}{\mbox{MeV}}
\newcommand{\bc}{\begin{center}}
\newcommand{\ec}{\end{center}}
\newcommand{\neea}{\nonumber\eea}
\newcommand{\lsm}{$L\sigma M\,\,$}
\newcommand{\bul}{$\bullet$}
\newcommand{\ed}{\end{document}}

\title{\LARGE
Light-quark dynamics\thanks{Lectures given at the
  41.\ Internationale Universit\"atswochen f\"ur Theoretische Physik,
\textit{Flavour Physics}, Schladming, Austria, February
22--28, 2003. To appear in the Proceedings.}}
\author{J\"urg Gasser 
\thanks{e-mail: gasser@itp.unibe.ch}\\
\small Institute for Theoretical Physics, University of Bern \\
\small Sidlerstr. 5, CH-3012 Bern, Switzerland}

\maketitle
\begin{abstract}
I present introductory lectures on the use of effective 
field theories in the low-energy regime of QCD.
\end{abstract}

\vskip8cm

{\footnotesize{\begin{tabular}{ll}
{\bf{Pacs:}}$\!\!\!\!$& 11.30.Rd, 12.39.Fe, 13.75.Lb, 
11.80.Et\\
{\bf{Keywords:}}$\!\!\!\!$& Chiral symmetries, 
Chiral Perturbation Theory, Meson-meson interactions,\\&
Pion-pion scattering, Roy equations, Partial wave analysis
\end{tabular}} }

\clearpage
\tableofcontents
\newpage
\setcounter{equation}{0}
\section{Introduction}
At low energies $E \ll M_{W}$, the interactions of
leptons and hadrons are
described by QCD + QED up to corrections of order $(E/M_{W})$.
 If we disregard
the electromagnetic interactions, we are
 left with QCD that
contains only a few parameters: the
renormalization group invariant scale $\Lambda$ and the running
quark masses $m_{u},m_{d},m_{s} \ldots$.
The quark masses $m_{u},m_{d}$ and $m_{s}$ are small on a 
typical hadronic
scale like the mass of the rho or of the proton.
It makes therefore  sense to consider the limit where these
masses are set equal to
zero (chiral limit). The remaining
quarks $c,b, \ldots$ are not
light: although one may of course study the theoretical
 limit in which these
 masses also vanish, it
does not seem to be possible to recover the actual mass 
values by an expansion around that limiting
case. At low energies, a better approximation is obtained
 if the quarks $c, b,
\ldots$ are instead
treated as infinitely heavy. In this limit, the degrees of 
freedom associated
with these quarks freeze and may be ignored in the effective 
low
 energy theory.

In the chiral limit, QCD  contains therefore 
only one parameter,
the scale $\Lambda$. The mass of the
proton is a pure number multiplying $\Lambda$, and likewise for
all the other states of the
theory --  the numbers $M_ {\rho} /M_{p}, M_{\Delta} /M_{p},
\ldots$ are determined in a parameter free
manner. In this sense, the chiral limit of QCD may be called a
 theory without any adjustable
parameters: QCD is of course unable to predict the value of 
$M_{p}$ in GeV
units, but it determines
all dimensionless hadronic quantities in a parameter free
 manner. The elastic cross section for $pp$
scattering e.g. is some fixed function of the variables
 $s/M_{p}^{2}$
\mbox{and}
$t/M_{p}^{2}$ , multiplying the square of the Compton
 wavelength of the proton.

It is unfortunately very difficult
 to really {\em
calculate \/} masses, cross sections and decay amplitudes 
in this beautiful
theory, because the lagrangian of QCD is 
formulated in terms of 
quark and
gluon fields which do not create asymptotically observed
particles.  Several methods have
therefore been
devised  in the past to cope with this problem in different
 regimes of the
energy scale:

{\em i) Processes at high energies. \/} At high energies, 
the effective
coupling constant ${\alpha}_\mathrm{QCD}$ becomes small, and 
conventional
perturbation theory in ${\alpha}_\mathrm{QCD}$ is applicable.

{\em ii) Lattice calculations.\/} This is the only method 
known today which
leads directly
from the QCD lagrangian to the mass spectrum, decay matrix
elements, scattering lengths
 etc.
 On the other hand, the CPU time needed for full
fledged QCD
calculations is enormous, and I believe that one may still
 have
to wait a long time
before this program achieves the accuracy one is aiming at in
the framework of effective field theory.

{\em iii) Chiral perturbation theory\/} (ChPT). This method
exploits the symmetry of the QCD
lagrangian and its ground state: one solves in a perturbative
manner the constraints
imposed by chiral symmetry and unitarity by expanding the 
Green functions
in powers of  the external momenta and of  the quark
masses $m_u,m_d$ and $ m_s.$
 To illustrate the idea, consider the
process
$\pi^+(p_1)\pi^-(p_2)\rightarrow\pi^0(p_3)\pi^0(p_4)$.
 Chiral
symmetry implies that the corresponding 
scattering amplitude has the following form near
threshold,
\bea\label{eq:tcharged}
T=\frac{M_\pi^2-s}{F_\pi^2}
+O(p^4)
\sem s=(p_1+p_2)^2\co
\eea
where $F_\pi=92.4$ MeV is the pion decay constant, 
 and  $M_\pi$ denotes
the pion mass.
This result
 is due to Weinberg~\cite{wein66}, who used
current algebra
and PCAC to analyse the Ward identities for the four-point
functions of the  axial currents.
It  displays the first order term in a systematic
expansion
of the scattering amplitude  in powers of momenta and of quark
masses. This term  algebraically
dominates the remainder, denoted by the symbol $O(p^4)$,  for
sufficiently small energies and thus provides
an accurate parameterization of the full amplitude near 
threshold. As one goes away from threshold, 
the higher order terms
  come into
play. We will see   in the following  that ChPT is a
method that allows one to determine these corrections 
in a systematic manner. 

ChPT is a particular example of an 
{\em effective field theory} (EFT).
 The method is in use since about 20 years,
 and it was therefore not possible to provide a 
 detailed review  in my lectures -- 
 for a recent comprehensive introduction to ChPT, 
I refer the reader to~\cite{scherer}.
Instead, I discussed a few basic principles and applications,
 in the hope that students become interested in 
this fascinating topic and continue with their own studies 
and research projects.

The article is organized as follows. In section 2, 
the flavor symmetries of QCD  are discussed, 
and their Nambu-Goldstone
realization explained. In section 3, the Goldstone theorem is 
 stated and illustrated with the free scalar field, 
 with the linear sigma model (\lsm$\!$) and with QCD. 
In addition,
 the
interaction of the Goldstone bosons at low energy is
investigated. Section 4 contains a discussion of the 
effective field theory 
 of the \lsm and of QCD at low energy. In section 5 are
 illustrated
 some calculations with these EFT, and section 6 contains a
  detailed
 discussion  of the elastic $\pi\pi$ scattering 
amplitude in this framework.
 In section 7, it is shown how Roy equations may 
be used to
determine low-energy constants that appear in the
 calculation of the $\pi\pi$ 
scattering amplitude. A short outlook on other topics 
is given in section 8. 

\clearpage
\setcounter{equation}{0}
\section{QCD with two flavours}
\label{sec:qcdgb}
In this section, I discuss the flavour symmetries of QCD.
\subsection{Symmetry of the lagrangian}
The lagrangian of QCD is
\bea\label{eq:lqcd}
{\cal L}&=&-\frac{1}{2g^2}\langle G_{\mu\nu}G^{\mu\nu}
\rangle_c 
+{\cal L}_{ud}\scs
\eea
where
\bea
{\cal L}_{ud}&=&\bar{u}\!\not\!\!{D} u+
\bar{d}\!\not\!\!{D}d -m_u\bar{u}u -m_d \bar{d}d
\nonumber\\[2mm]
&=&(\bar{u}\,\,\bar{d})\biggl(\begin{tabular}{cc}
$\!\not\!\!{D}-m_u$&$0$\\
$0$&$\!\not\!\!{D}-m_d$
\end{tabular}\biggr) 
\biggl(\begin{tabular}{c}$u$\\$d$\end{tabular}\biggr)
\sem \!\not\!\!{D}=i\gamma^u(\partial_\mu-iG_\mu)\fs
\nonumber\eea
$G_{\mu\nu}$ denotes the field strength associated with the 
gluon field $G_\mu$,  and $\langle A \rangle_c$ stands for the 
color trace of the matrix $A$.

\begin{figure}[H]
\bc
\epsfig{figure=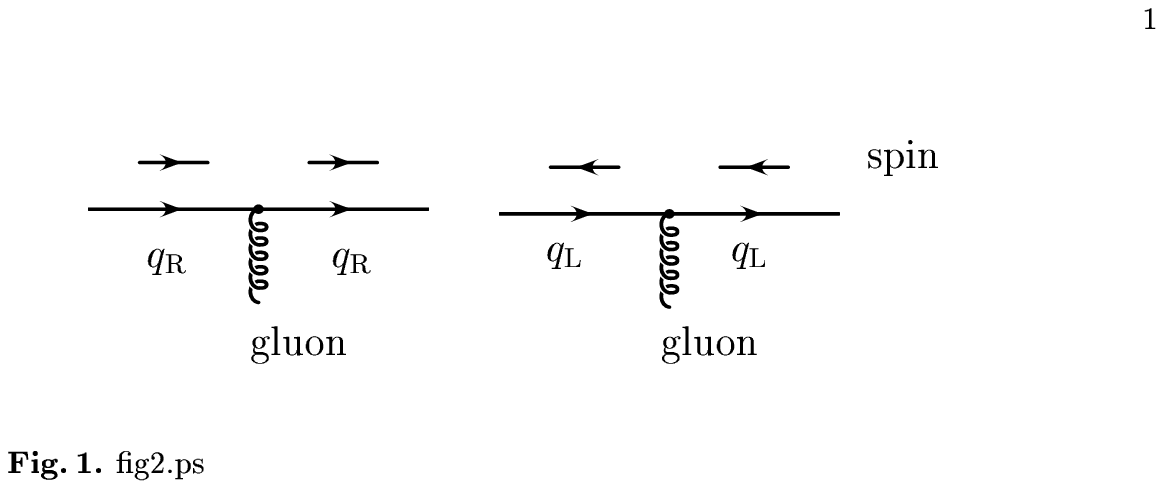,
height=2.5cm,
width=8cm,
bbllx=141,
bblly=634,
bburx=414,
bbury=727,
clip=
}
\caption{Gluon interactions do not change the helicity 
of the quarks}\label{fig:helicitygluons}
\ec
\end{figure}

\begin{figure}[H]
\bc
\epsfig{figure=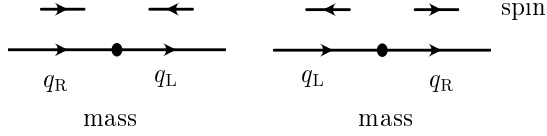,
height=2.5cm,
width=8cm,
bbllx=118,
bblly=655,
bburx=392,
bbury=739,
clip=
}
\caption{The mass terms  change the 
helicity of the quarks}\label{fig:helicitymass}
\ec
\end{figure}

It is useful to introduce left- and right-handed spinors,
\bea
u_{L}&=& \frac{1}{2}(1-\gamma_5)u\scs
u_{R}=\frac{1}{2}(1+\gamma_5)u\nonumber\scs
\eea
\bea
{\cal L}_{ud}
&=&(\bar{u}_L\,\,\bar{d}_L)\biggl(\begin{tabular}{cc}
$\!\not\!\!{D}$&$0$\\
$0$&$\!\not\!\!{D}$
\end{tabular}\biggr) 
\biggl(\begin{tabular}{c}$u_L$\\$d_L$\end{tabular}\biggr)
 -
(\bar{u}_L\,\,\bar{d}_L)\biggl(\begin{tabular}{cc}
$m_u$&$0$\\
$0$&$m_d$
\end{tabular}\biggr) 
\biggl(\begin{tabular}{c}$u_R$\\$d_R$\end{tabular}\biggr)
+L\leftrightarrow R\fs
\neea
QCD makes sense for any value of the quark masses. For 
$m_u=m_d=0$, the lagrangian (\ref{eq:lqcd}) 
is invariant under $U(2)$ rotations of the left- and 
right-handed fields,
\bea\label{eq:u2trafo}
\biggl(\begin{tabular}{c}$u_I$\\$d_I$\end{tabular}\biggr )
\Rightarrow V_I
\biggl(\begin{tabular}{c}$u_I$\\$d_I$\end{tabular}\biggr )
\sem 
V_I\in U(2)\scs I=L,R\fs
\eea
In other words, gluon interactions do not 
change the helicity of the
quarks, see Fig.~\ref{fig:helicitygluons}. On the other hand, 
the terms proportional to the quark masses are not invariant 
under the transformations (\ref{eq:u2trafo}), 
see Fig.~\ref{fig:helicitymass}.  

According to the theorem of E. Noether, 
 there is one conserved current for each 
continuous parameter in the symmetry group. As
the group $U(2)$ has  four real parameters, one expects 
eight conserved currents.
 However, due to quantum effects, one of these currents 
is not conserved, as a result of which there are only 
seven conserved currents in the limit of vanishing quark 
masses,
\bea\label{eq:currents}
L_\mu^a&=&\bar q_L\gamma_\mu\frac{\tau^a}{2}q_L\scs
R_\mu^a=\bar q_R\gamma_\mu\frac{\tau^a}{2}q_R\sem
a=1,2,3\nn
V_\mu&=&\bar q\gamma_\mu q\sem
q=\biggl (\begin{tabular}{c}$u$\\$d$\end{tabular}\biggr)\fs
\eea
The world at 
\bea
m_u=m_d=0\neea
is called {\em the chiral limit of QCD}, and 
the above statements are summarized as:
{\em In the chiral limit,
 ${\cal L}_\mathrm{QCD}$ 
is symmetric under global 
{$SU(2)_L\times SU(2)_R\times U(1)_V$} transformations.
The corresponding 7 Noether currents (\ref{eq:currents})
 are conserved.}

\subsection{Symmetry of the ground state}

 It is useful to introduce in addition 
the vector and axial currents
\bea
V^{\mu a} &=& 
\bar q \gamma^\mu \frac{\tau^a}{2}q =L^{\mu a}+R^{\mu a}\scs 
\nonumber\\
A^{\mu a} &=& 
\bar q \gamma^\mu \gamma_5\frac{\tau^a}{2}q = 
R^{\mu a}- L^{\mu a}\sem a=1,2,3 \fs
\nonumber\eea
The corresponding 6 axial and vector 
charges $Q_{A,V}^a$  are conserved and 
commute with the hamiltonian
 $H_0=H_\mathrm{QCD}|_{m_u=m_d=0}$,
\bea\label{eq:conservedcharges}
[H_0,Q_V^a]=[H_0,Q_A^a]=0\sem a=1,2,3\fs
\eea
Consider now  eigenstates of $H_0$,
\bea
H_0|\psi\rangle =E|\psi\rangle\fs
\nonumber\eea
Then the states
 $Q_A^a|\psi\rangle\,\, {\mbox{and}}\,\, 
Q_V^a|\psi\rangle$
 have the same energy $E$, but carry opposite parity.
 On the other hand, there is no trace \cite{pdg}   of 
such a symmetry in nature. The resolution of the paradox 
has been provided by Nambu and Lasinio back in 1960 
\cite{nambu}: whereas the vacuum is annihilated by 
the vector charges, it is not invariant under the action 
of the axial charges,
\bea\label{eq:vacannihilation}
Q_V^a|0\ra=0\scs Q_A^a|0\rangle\neq 0\fs
\eea
There are two important  consequences of this 
assumption:
\begin{itemize}
\item[i)] 
The spectrum of $H_0$ contains  three massless, 
pseudoscalar particles (Goldstone bosons (GB); 
Goldstone~\cite{goldstone}). We will see more of this in the 
 following section.
\item[ii)]
The axial charges $Q_A^a$, acting on any state in the Hilbert 
space, generate Goldstone bosons,
\bea
Q_A^a|\psi \rangle = |\psi, G_1,\ldots, G_N,\ldots\rangle\fs
\neea
These are not one-particle states, and are therefore 
 not listed in PDG, 
and there is therefore no contradiction anymore.
\end{itemize}
A theory with (\ref{eq:conservedcharges}),
 (\ref{eq:vacannihilation}) is called 
{\em spontaneously broken}: the
symmetry of the hamiltonian is not the same as 
the symmetry of the ground state.

Where are the three massless, pseudoscalar states?
 The three pions $\pi^\pm,\pi^0$ are the lightest hadrons.
 They are not massless, because the quark masses are 
not zero in the real world~\cite{leutwylerratios}:
\bea\label{eq:quarkmasses}
m_u&\simeq& 5\, \MeV\co\nn
m_d&\simeq& 9\, \MeV\fs
\eea
In the following, we assume that the flavour symmetry  
of QCD is spontaneously 
broken to the diagonal subgroup,
\bc
\framebox{$SU(2)_L\times SU(2)_R \rightarrow SU(2)_V$}
\ec
and work out the consequences for the {\em interactions}
 between the Goldstone bosons.

\underline{Remark:} Vafa and Witten~\cite{vafa} have 
shown that -- modulo highly plausible assumptions --
 the vector symmetry $SU(2)_V$ is not spontaneously 
broken.

\subsection{A remark on isospin symmetry}
Even if the quarks are not massless, the  QCD lagrangian 
 has a residual symmetry at $m_u=m_d$: it is 
invariant under the 
transformations (\ref{eq:u2trafo}) with $V_R=V_L\in SU(2)$.
 This symmetry is called {\em  isospin symmetry}. We know 
from textbooks that isospin symmetry violations
in the strong interactions are 
small\footnote{There are two
 sources of isospin violations: those due to 
electromagnetic interactions, and those due to the 
difference in the up and down quark masses}. On the other 
hand, 
 according to (\ref{eq:quarkmasses}), one has
\bea
m_d/m_u\simeq 1.8\fs
 \eea
How can then isospin be a good symmetry if the quark 
masses differ so much? Consider the neutral and the 
charged pions: is it so  that their masses differ by
\bea
(M_{\pi^+}^2-M_{\pi^0}^2)/M_{\pi^0}^2\simeq
\frac{m_d-m_u}{m_d+m_u}\simeq 0.3\,\,?
\neea
The answer is no: one has
\bea
M_{\pi^+}^2=(m_u+m_d)B+\cdots\co\nn
M_{\pi^0}^2=(m_u+m_d)B+\cdots\co
\neea
where the ellipses denote higher order terms in the 
quark mass expansion. The neutral and the charged pion 
have the same leading term, the quark mass difference 
shows up only in the quadratic piece,
\bea
M_{\pi^+}^2-M_{\pi^0}^2=O[(m_u-m_d)^2]\fs
\neea
The perturbation due to the quark masses can be written as
\bea
m_u\bar u u +m_d \bar d d&=&\frac{1}{2}
(m_u+m_d)(\bar u u+\bar d d)\nn&+&\frac{1}{2}(m_u-m_d)
(\bar u u -\bar d d)\fs
\neea
Isospin is a good symmetry, not because $(m_d-m_u)/(m_d+m_u)$ 
is small, but because the matrix elements of the operator 
$\frac{1}{2}(m_d-m_u)(\bar u u -\bar d d)$
 are small with respect to the hadron masses. The bulk part 
in the pion mass
 difference is generated by electromagnetic interactions.

\setcounter{equation}{0}
\section{Goldstone bosons}
In this section, I discuss the Goldstone theorem, 
illustrate it with several examples and consider the 
interaction of Goldstone bosons at low energy.

\subsection{The Goldstone theorem}
We consider a  quantum field theory  which has the 
following properties: 
\begin{itemize}
\item[i)] There is  a conserved current 
(i.e., an object that
  transforms as a four-vector under proper Lorentz 
transformations),
\bea
A_\mu(x)\sem \, \partial^\mu A_\mu=0.
\neea
\item[ii)]
There is an operator $\Phi(x)$ such that
\bea\label{eq:fieldphi}
\langle 0|[Q,\Phi]|0\rangle\neq 0\sem Q=\int d^3x 
A_0(x^0,\vec{x}).
\eea
\end{itemize}
Then the Goldstone theorem~\cite{goldstone} applies:
\begin{enumerate}
\item
 There exists a massless particle in the theory,
\bea
|\pi(\mathbf{p})\rangle\scs p^2=0\fs
\neea
\item
 The current $A_\mu$ couples to the massless state,
\bea
\langle 0|A_\mu(0)|\pi(\mathbf{p})\rangle =ip_\mu F\neq 0\fs
\neea
\end{enumerate}
From the condition
(\ref{eq:fieldphi}), it is seen that the charge $Q$ does not
annihilate the vacuum.
\subsection{The free scalar field}
We begin with a very simple example, the free, massless 
 scalar field.
The lagrangian is given by
\bea
{\cal L}_0&\!=&\!\frac{1}{2}\partial_\mu\phi\partial^\mu\phi\fs
\neea
For the current $A_\mu$, we  take 
\bea
  A_\mu\!=\!\partial_\mu\phi\fs
\neea
This current is conserved, because $\phi$ is a free field.
 Consider now $\Phi=\phi$. From the canonical 
commutation relations,  it follows that the condition
 (\ref{eq:fieldphi}) is satisfied. Therefore, 
the Goldstone theorem applies. Indeed, we can easily check
 directly:
\bea
\bullet\,\,\phi\,\, \mbox{generates massless states}\co
\neea
and
\bea
\bullet 
\,\,\langle 0|A_\mu(0)|\pi\rangle=-ip_\mu\neq 0\fs
\neea

\subsection{The linear sigma model}
We consider the linear sigma model (\lsm$\!$), because it 
allows one to illustrate many features of  effective 
field theories. At the same time, it serves as a model with
spontaneous symmetry breaking. The lagrangian is
\bea\label{eq:laghiggs}
{\cal L}_\sigma=\frac{1}{2}\partial_\mu\vec{\phi}
\cdot \,\partial^\mu\vec{\phi}-\frac{g}{4}(\vec{\phi}^{2}
-v^2)^2\co
\eea
where
$\vec{\phi}=(\phi^0,\phi^1,\phi^2,\phi^3)$
denotes four real fields, and
$\vec{\phi}^{2}=\phi^k\phi^k$ 
[repeated indices are summed over in the absence of a 
summation symbol]. In the 
following, we assume that
\bea
v^2 > 0\co
\neea
and discuss
\begin{enumerate}
\item[\bul]
the symmetry properties of ${\cal L}_\sigma $
\item[\bul]
spontaneous symmetry breakdown
\item[\bul]
Goldstone bosons
\item[\bul]
quantization
\item[\bul]
Goldstone boson scattering
\end{enumerate}
{\bf Symmetry properties}
Here, we consider the classical theory and observe that
${\cal L}_\sigma $ is invariant under four-dimensional 
rotations of the vector $\vec{\phi}$,
\bea
\phi^i\rightarrow R^{ik}\phi^k\scs R\in O(4)\fs
\neea
The matrices $R$ can be parametrized 
in terms of six real parameters.
Let us consider infinitesimal rotations
\bea
R=1+\varepsilon+O(\varepsilon^2)\fs
\neea
Because $R$ is an orthogonal matrix, $\varepsilon$ is 
antisymmetric, $\varepsilon+\varepsilon^T=0$. Every real and 
antisymmetric four by four matrix can be expanded 
in terms of six generators,
\bea
\varepsilon=\sum_{i=1}^3\biggl(c_i\varepsilon_V^i+d_i
\varepsilon_A^i\biggr)\co
\neea
where $c_i,d_i$ are 6 real parameters. The 
generators 
\begin{center}
\epsfig{file=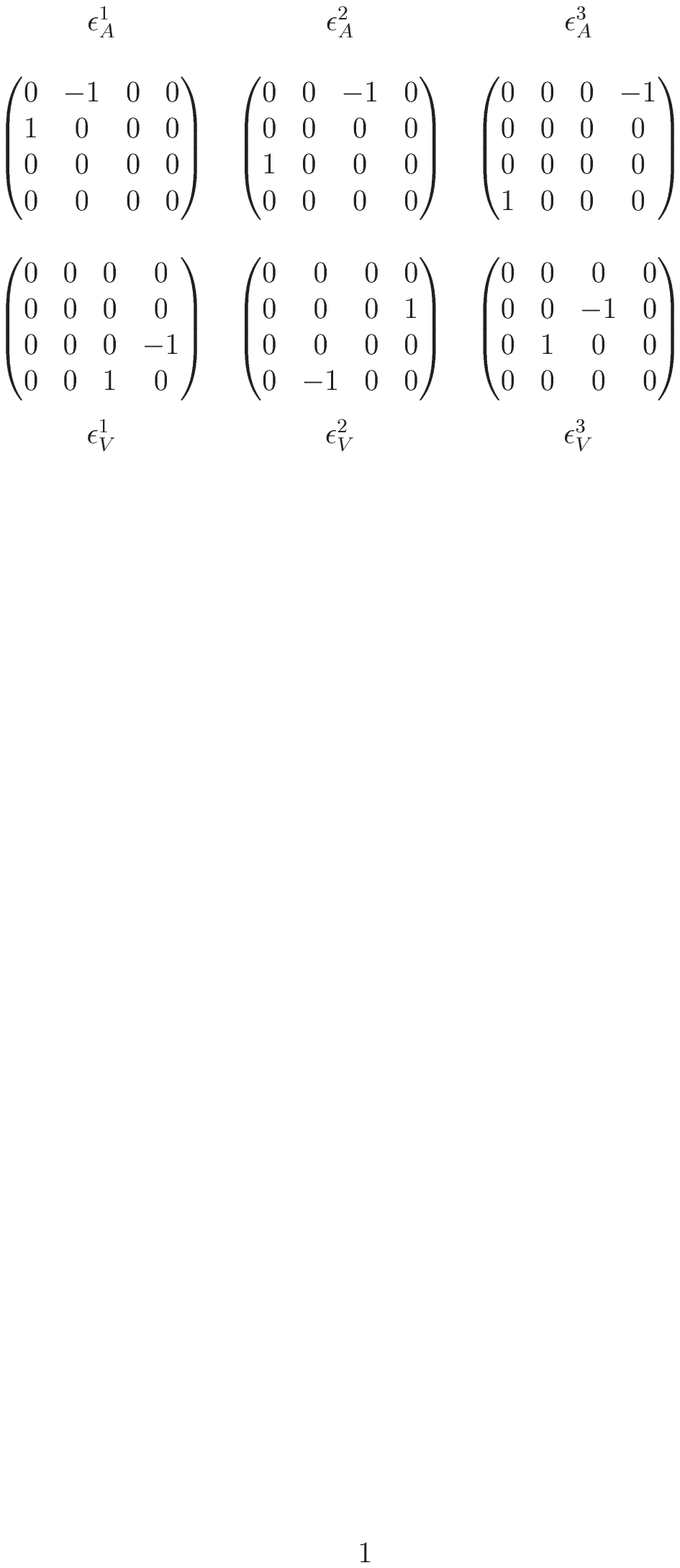,width=8cm,
bbllx=108,bblly=520,bburx=440,bbury=720,clip=}
\end{center}
 satisfy the commutation relations

$[ \varepsilon_V^a, \varepsilon_V^b ]
= \varepsilon^{abc}\varepsilon_V^c\co$

$[ \varepsilon_V^a, \varepsilon_A^b ]
= \varepsilon^{abc}\varepsilon_A^c\co$

$[ \varepsilon_A^a, \varepsilon_A^b ]
=\varepsilon^{abc}\varepsilon_V^c\co$

\noindent
with $\varepsilon^{123}=1$, cycl. The linear combinations
\bea
Q_L^a=\frac{1}{2}(\varepsilon_V^a-\varepsilon_A^a)\scs
Q_R^a=\frac{1}{2}(\varepsilon_V^a+\varepsilon_A^a)\co
\neea
generate two commuting $SU(2)$ Lie-algebras 
(up to a factor $i$),
\bea
\begin{tabular}{ll}
$[Q_I^a,Q_I^b]=\varepsilon^{abc}Q_I^c$&$\sem I=L,R\co$\\
$[Q_L^a,Q_R^b]=0 \fs$
\end{tabular}
\neea
In other words, the lagrangian ${\cal L}_\sigma $ has a 
$SU(2)_L\times SU(2)_R$ symmetry. As a result of this, 
 there are six conserved Noether currents,  which we take to be
\bea
V_\mu^a&=&\varepsilon^{abc}\phi^b\partial_\mu\phi^c\co\nn
A_\mu^a&=&-\phi^0\stackrel{\leftrightarrow}
\partial_\mu\phi^a\sem a=1,2,3\fs
\neea

\subsubsection*{Spontaneous symmetry breaking}
\begin{figure}[H]
\bc
\epsfig{figure=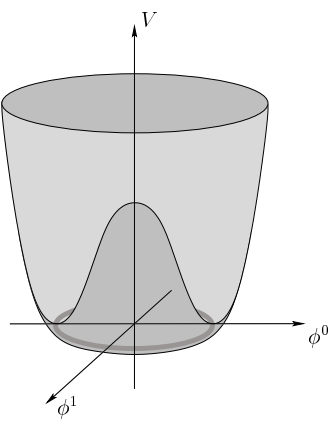,
width=6cm,
bbllx=203,
bblly=504,
bburx=444,
bbury=742,
clip=
}
\caption{
Potential for $O(2)$}\label{fig:potential}
\ec
\end{figure}

The potential $V={g}(\vec{\phi}^{\,\,2}-v^2)^2/4$
is extremal at $\phi=0$ and at $\vec{\phi}^{\,\,2}=v^2$. 
 The latter configuration corresponds to a global minimum. 
The vector
\bea
\vec{\phi}_G=(v,\vec{0})\co 
\neea
which realizes this global minimum,
 is only invariant under the subgroup $H=O(3)$ 
(with generators $\varepsilon_V^a$), see 
figure \ref{fig:potential} for the case where the 
symmetry group is  $O(2)$.
The number of generators that do not leave invariant 
$\vec{\phi}_G$ is $n_G-n_H=3$, where
\bea
n_G:&&\mbox{$\#$ of parameters in O(4)}\co\nn
n_H:&&\mbox{$\#$ of parameters in O(3)}\fs
\neea
Therefore,  one expects three Goldstone bosons in the 
spectrum of the theory.

\subsubsection*{Goldstone bosons}
In order to identify the Goldstone bosons, we consider 
fluctuations around the configuration $\vec{\phi}_G$, 
and write
\bea
\vec{\phi}=(v+\varphi_0,\vec{\pi})\sem \vec{\pi}=
(\pi^1,\pi^2,\pi^3)\co
\neea
where we have introduced the pion fields $\vec{\pi}$. 
In terms of the new fields,  the lagrangian becomes
\bea\label{eq:laghiggs1}
{\cal
  L}_\sigma &=&\frac{1}{2}[\partial_\mu\varphi_0\partial^\mu
\varphi_0-2gv^2\varphi_0^2] +
\frac{1}{2}\partial_\mu\vec{\pi}
\cdot\partial^\mu\vec{\pi}\nn
&& -gv\varphi_0\,(\varphi_0^2+\vec{\pi}^{\,2})
-\frac{g}{4}
(\varphi_0^2
+\vec{\pi}^{\,\,2})^2\fs
\eea
The kinetic term shows that there is indeed
 one massive field $\varphi_0$, with mass  $m=\sqrt{2g}v$, 
together with three  massless fields $\vec{\pi}$.

\subsubsection*{Quantization}
One evaluates Green functions generated by the 
la\-gran\-gian
(\ref{eq:laghiggs1}) in the standard manner. 
At tree level, one has one massive and three 
massless fields, as in the classical theory.
 Evaluating loops, one finds that $\varphi_0$ picks up a vacuum
 expectation value at order $\hbar$. One shifts this 
field again,
 such that the new field has a vanishing one-point 
Green function, and finds that the remaining three particles 
stay massless also in the loop expansion.

\subsubsection*{Goldstone boson scattering}
\begin{figure}[H]
\bc
\epsfig{figure=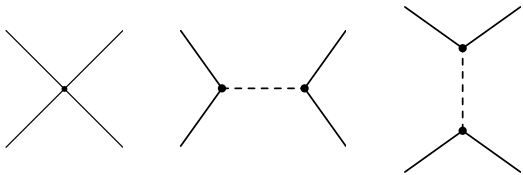,
width=8cm,
bbllx=158,
bblly=640,
bburx=423,
bbury=740,
clip=
}
\caption{Goldstone boson scattering  with the lagrangian
  (\protect{\ref{eq:laghiggs1}}).  {\it Solid (dashed)}
lines stand for 
massless (massive) particles}\label{fig:tree}
\ec
\end{figure}

Finally, we consider GB scattering at tree level, 
with the lagrangian (\ref{eq:laghiggs1}). In particular, 
we consider the process
\bea
\pi^1(p_1)\pi^2(p_2)\rightarrow\pi^3(p_3)\pi^4(p_4)\co
\neea
where $\pi^1$ denotes the GB number one, etc.
 The relevant diagrams  are displayed in Fig.~\ref{fig:tree}. 
The scattering matrix element has the structure
\bea\label{eq:Apipi}
T^{kl;ij}&=&\delta^{ij}\delta^{kl}A(s,t,u)+\mbox{cycl.}\co\nn
&&s=(p_1+p_2)^2,t=(p_3-p_1)^2,u=(p_4-p_1)^2\fs
\eea
(The Kronecker symbols occur, because the lagrangian 
is invariant under an $O(3)$ 
 rotation of the pion fields $\vec{\pi})$.
 The invariant amplitude is
\bea\label{eq:pipisigma}
A(s,t,u)=\frac{4g^2v^2}{m^2-s}-2g\fs
\eea
At small momenta, the constant terms cancel out,
\bea\label{eq:pipilinsig}
A(s,t,u)=\frac{s}{v^2}+O(s^2)\fs
\eea
In other words, the interaction between the GB vanishes at
 zero momenta. The constant $v$ 
is related to the matrix elements of the axial 
current,
\bea
\langle 0|A_\mu^a(0)|\pi^b(p)\rangle=i\delta^{ab}vp_\mu\fs
\neea

\subsection{QCD with two flavors}
We now go back to the discussion of the Goldstone 
theorem in the framework of QCD. In section 
\ref{sec:qcdgb}, we noted that the three 
axial currents $A_\mu^a$ are conserved in the case of 
vanishing quark masses. For the field $\Phi$ in 
(\ref{eq:fieldphi}), we choose
$\Phi=\bar q\gamma_5\tau^a q$ (three fields, one for each a). 
Applying canonical commutation
relations, one finds that  
\bea
[Q_A^a,\Phi^b]=-\delta^{ab}(\bar u u + \bar d d)\sem 
a,b=1,2,3\fs
\neea
Provided that the vacuum expectation 
value of the quark bilinear 
is different from zero, the Goldstone theorem applies:
\bea
&&\langle 0|\bar u u|0\rangle \neq 0 
\Rightarrow {\mbox{3  Goldstone bosons}}\fs
\neea
(The vacuum expectation value of $\bar u u$ is equal to the 
one of $\bar d d$ by isospin symmetry.) Lattice 
calculations support the conjecture that 
$\langle 0|\bar u u|0\rangle$ is different from zero. 
Later in these lectures, we shall see  that data on 
$K_{e4}$ decays do so as well.

\begin{figure}[H]
\begin{center}
\epsfig{figure=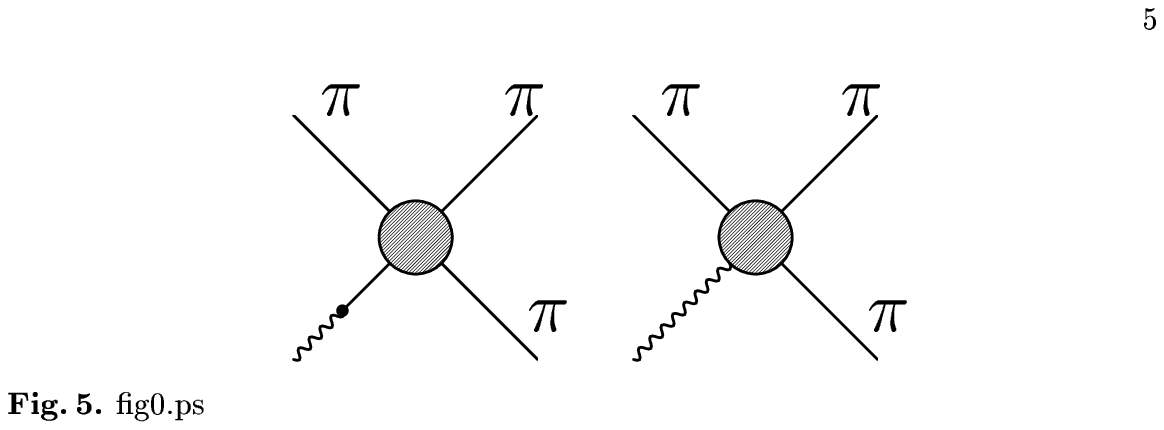,
width=6cm,
bbllx=199,
bblly=640,
bburx=423,
bbury=740,
clip=
}
\caption{Singular and non singular contributions to the 
matrix element (\protect{\ref{eq:gpipi}}) of the axial 
current. The {\it wavy line} denotes the axial current, the 
{\it solid lines} the pions}\label{fig:apipi}
\end{center}
\end{figure}

How does one evaluate GB scattering in QCD?
This is a very complicated affair: 
 the QCD lagrangian contains quark and gluon fields, 
not pion fields. On the other hand, if QCD is spontaneously 
broken in the manner just discussed, the 
Goldstone theorem guarantees that the axial 
current can be used as an interpolating field for 
the pion~\cite{haag}.
Let us consider therefore the matrix element
\bea\label{eq:gpipi}
G_\mu(p_4,p_3;p_1)=
\langle \pi (p_3)\pi (p_4){\mbox{out}}|A_\mu(0)
|\pi(p_1)\rangle\co
\eea
where I have suppressed all isospin indices. This 
matrix element has two parts: one, where the axial current 
generates a
pion pole, and a second one, which is free from one-particle
singularities, see Fig.~\ref{fig:apipi}.
According to the LSZ reduction formula~\cite{lsz}, the 
quantity $G_\mu$ has the structure
\bea\label{eq:gbscattering}
G_\mu=\frac{Fq_\mu}{q^2}T(p_3,p_4;p_1) +R_\mu\co
\eea
where $T$ denotes the elastic $\pi\pi$ scattering 
matrix element, and where $F$ is the pion decay 
constant, 
\bea
\la0|A_\mu(0)|\pi(p)\ra=ip_\mu F\fs
\neea
The remainder 
$R_\mu$ is non singular when $q^\mu=(p_3+p_4-p_1)^\mu$ 
is sent to zero.
We contract both sides in (\ref{eq:gbscattering}) 
with $q^\mu$. 
Because the axial current is
conserved, the left-hand side vanishes, and therefore
\bea
FT(p_3,p_4;p_1)+q^\mu R_\mu=0\fs
\neea
One concludes that 
the scattering matrix element vanishes at $q^\mu=0$ - 
this was easy to prove! We find again that the GB do 
not interact at vanishing momenta. 
One can even go further: as already mentioned in the 
introduction,  Weinberg determined
in 1966~\cite{wein66} -- using current algebra -- 
the leading term of the $\pi\pi$ 
scattering amplitude in a systematic 
expansion of the momenta and of
the pion mass.
\setcounter{equation}{0}
\section{Effective field theories }
\begin{figure}[H]
\bc
\epsfig{figure=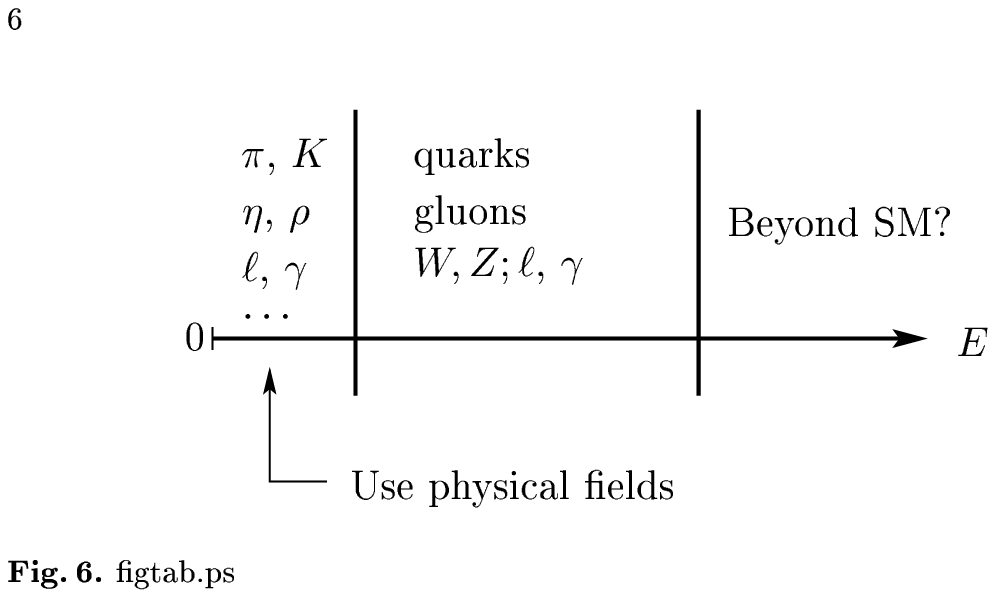,
height=4cm,
bbllx=167,
bblly=598,
bburx=440,
bbury=740,
clip=
}
\caption{The Standard Model at various 
energy scales. At low energy,
 the effective theory is formulated in terms of physical 
fields}
\label{fig:effective}
\ec
\end{figure}

As we have just seen, it is possible to get a  great deal of
information about the interactions of GB in QCD 
without actually solving the theory. 
{\em Effective field theories} (EFT) provide the proper 
framework 
 to perform detailed calculations in a systematic 
manner. EFT are
 valid in a restricted energy region,
describing there an underlying theory that is valid on 
a wider energy scale. The situation for the Standard Model 
 is 
 illustrated in  Fig.~\ref{fig:effective}.

Effective field theories  are in particular useful, when
 a full calculation is not yet possible, 
as is the case with the Standard Model at low energies.
 One sets up  an EFT with the same symmetry properties 
 -- in case of the Standard Model, this effective theory is 
 called 
{\em Chiral perturbation theory}.
A second possibility for the use of EFT occurs in case that the
 calculations can be done in the underlying theory, 
but are very complicated. An example is provided by the 
calculation of bound states in the framework of QED, 
which can be performed, in principle,  at any
 desired order by use of the Bethe-Salpeter equations. On 
the other
 hand, as Caswell and Lepage have shown~\cite{caswell},
 the use of EFT makes life
 very much simpler also in this case,  because the 
electrons and
 positrons are moving very non-relativistically in the 
atoms, and a relativistic calculation is not really needed.
Further applications concern the case where the 
underlying theory is
 not known -- one builds the effective theory in terms of 
the light
 fields and attempts to determine from experiment the unknown 
coupling constants that occur in there.

In the following, I discuss the low-energy EFT  for the \lsm
and for QCD.

\subsection{Linear sigma model at low energy}

We have seen in the last section that the  $O(4)$ 
version of the \lsm in its 
broken phase develops three Goldstone bosons, 
which interact weekly 
 at low energy (we have proven this at tree level only).

In the following, we construct an effective theory that 
contains only pions and 
their interactions - the sigma particle is removed 
from the theory. This EFT is constructed 
in such a manner that the Green functions with pion 
fields, evaluated in the \lsm at low energies, are recovered 
by the EFT. How  can this be achieved?

As a first step, one constructs an EFT which 
reproduces all {\em tree graphs} at low energies, 
to any order in the low-energy expansion: as is seen from 
the explicit expression (\ref{eq:pipisigma}), the 
scattering matrix element contains arbitrarily high 
powers in the momenta even at tree level. The procedure to
construct the effective lagrangian is described in 
detail in the article by Nyffeler and 
Schenk \cite{nyffeler schenk} - 
I do not repeat the argument here and refer the 
interested reader instead to this work.
The result can be written in various equivalent forms. 
Here, I use
\bea
{\cal L}_\mathrm{eff}^\sigma &=&{\cal L}_2^\sigma+
{\cal L}_4^\sigma+\cdots\fs
\neea
The lagrangians ${\cal L}_n^\sigma$ contain $n$ 
derivatives of the pion fields. These derivatives 
become momenta in Fourier space: the effective lagrangian 
provides a momentum expansion of the amplitudes. 
Explicitly, one has
for the leading term
\bea\label{eq:lagefflsm}
{\cal L}_2^\sigma&=&\frac{v^2}{4}\la \partial_\mu U \partial 
^\mu U^\dagger\ra\co
\eea
where $U$ is a $2\times 2$ unitary matrix,
\bea
U&=&\sigma\cdot{\bf 1}_{2\times 2} +\frac{i}{v}
\tau^k\pi^k\,\sem\,
\sigma^2+\frac{\vec{\pi}^2}{v^2}=1\co
\vec{\pi}=(\pi^1,\pi^2,\pi^3)\fs
 \neea
The symbol $\la A\ra$ denotes the trace of 
the matrix $A$, and $\tau^k$ are the Pauli matrices.
In order to illustrate the structure of this lagrangian,
 we expand it in terms of the pion fields:
\bea
{\cal L}_2^\sigma=\frac{1}{2}\partial_\mu
{\vec{\pi}}\cdot
\partial^\mu
{\vec{\pi}}
+\frac{1}{8v^2}\partial_\mu
{\vec{\pi}}^ 2
\partial^\mu
{\vec{\pi}}^2
+O({\vec{\pi}}^6)\fs
\neea
The lagrangians ${\cal L}_{4,6,8\ldots}^\sigma$ have 
a similar  structure 
\cite{nyffeler schenk}.

\vskip.5cm

{\textbf{Comments}}

\begin{itemize}
\item
Only pions occur in the effective theory, the heavy 
particle has disappeared.
\item
It is easy to calculate the $\pi\pi$ scattering 
amplitude at tree level with ${\cal L}_2^\sigma$ - 
only one diagram remains 
to be calculated.
The result is identical to the first term on the 
right-hand side of (\ref{eq:pipilinsig}). 
 The higher order 
terms in the expansion of the amplitude 
(\ref{eq:pipisigma}) are generated by
 the tree graphs of ${\cal L}_{4,6,..}^\sigma$.
\item
The mass of the sigma particle is given by
 $m=\sqrt{2g}v$.
As a result of this, the interactions disappear formally 
in the large-mass limit. 
\item[5.]
Where has the $O(4)$ symmetry of the \lsm gone?
$
{\cal L}_2^\sigma$ is invariant under
\bea
U\rightarrow V_RUV_L^\dagger; \,\, V_{R,L}\in SU(2)\fs
\neea
Therefore, the effective theory has an 
$SU(2)_R\times SU(2)_L$ symmetry, as 
 the original theory.
\item[6.] 
How is the spontaneously  broken symmetry realized?

$U_G={\bf 1}_{2\times 2} $ is the ground state. 
It is only invariant under
\bea
U_G\rightarrow VU_GV^\dagger\fs
\neea
\end{itemize}
Therefore, the theory is spontaneously broken,
\bea
SU(2)_R\times SU(2)_L\Rightarrow SU(2)_V\co
\neea
and generates three Goldstone bosons.

 The  lagrangian ${\cal L}_\mathrm{eff}^\sigma$ 
reproduces the tree graphs of 
the \lsm - how about loops? Indeed, one may 
 calculate the  scattering amplitudes 
in the framework of the \lsm to any order in the 
loop expansion, perform a low-energy expansion 
of the result,  and finally construct an effective theory that 
reproduces  the result of this calculation, 
order by order in the low-energy expansion. 
The procedure is carried out to one loop 
in \cite{annals,nyffeler schenk}. 

\subsection{QCD at low energy}
\begin{figure}[H]
\begin{center}
\epsfig{figure=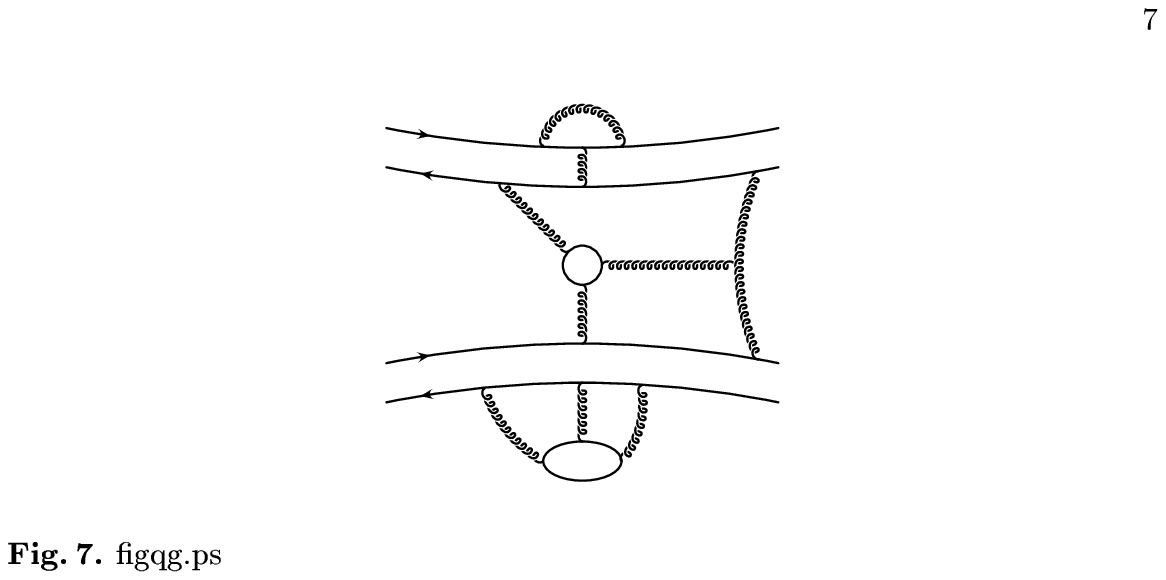,
width=6cm,
height=4cm,
bbllx=220,
bblly=600,
bburx=390,
bbury=730,
clip=
}
\caption{$\pi\pi$ scattering in QCD. Displayed is one of the
  infinitely many graphs that contribute in QCD, 
and which are all  equally important}\label{fig:pipiqcd}
\label{pipiqcd}
\end{center}
\end{figure}

\begin{figure}[H]
\begin{center}
\epsfig{figure=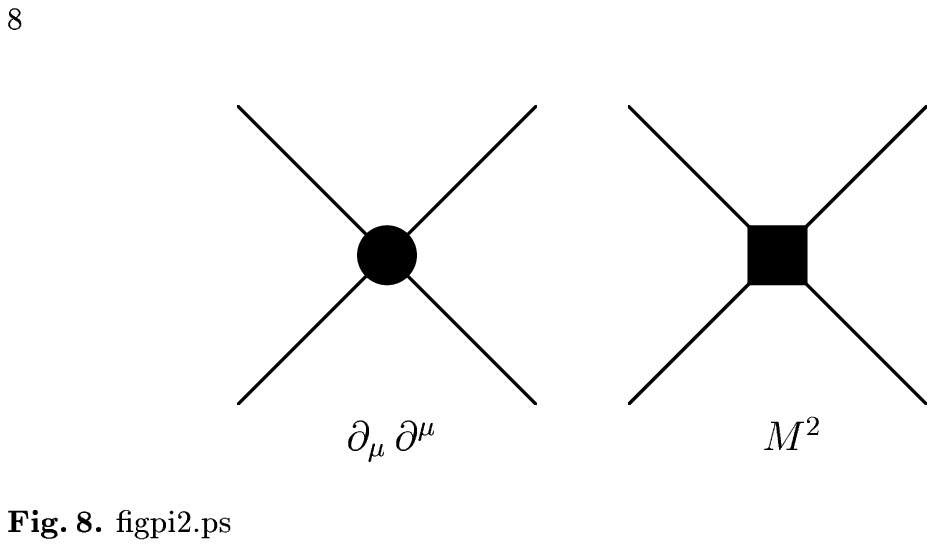,
height=3.5cm,
bbllx=182,
bblly=611,
bburx=432,
bbury=732,
clip=
}
\caption{Evaluating the $\pi\pi$ scattering amplitude 
with the  effective lagrangian
 ${\cal L}_2$  in ({\protect{\ref{eq:qcdleading}}}). 
Only the two graphs displayed contribute at leading order. 
The symbol $M^2$ is defined in \protect(\ref{eq:M2})}
\label{fig:pipieff}
\end{center}
\end{figure}

 The effective theory of QCD is formulated in  terms 
of asymptotic pion fields. As is the case for the 
\lsm$\!\!$, the effective theory 
consists of an infinite number of terms, with more and 
more derivatives~\cite{wein99,annals}:
\bea\label{eq:leffqcd}
{\cal L}_\mathrm{eff}={\cal L}_2+{\cal L}_4+{\cal L}_6
+\cdots\fs
\eea
All that goes into the construction of 
${\cal  L}_\mathrm{eff}$ are symmetry properties of QCD.  
The result for the leading term is
\bea\label{eq:qcdleading}
{\cal L}_2&=&\frac{F^2}{4}\la \partial_\mu U \partial 
^\mu U^\dagger+2B{\cal M}(U+U^\dagger)\ra\co
\eea
where the field $U$ is the same as above, and where
\bea
{\cal M}=\biggl(\begin{tabular}{ll}
$m_u$&0\\0&$m_d$\end{tabular}\biggr)
\neea
contains the quark masses $m_u,m_d$. A glance at 
(\ref{eq:lagefflsm}) shows that, at $m_u=m_d=0$, the 
leading order lagrangians in the \lsm and in QCD 
agree, provided that one sets $v=F$.
 The leading  term (\ref{eq:qcdleading}) 
contains the two constants 
$F$ and $B$ which are related to the pion decay constant 
and to the quark condensate, respectively \cite{annals},
\bea
\la 0|A_\mu^a(0)|\pi^b(p)\ra&=&ip_\mu F\delta^{ab}\co\nn
\la 0|\bar u u|0\ra_{|_{m_u=m_d=0}}&=&-F^2B\fs
\neea
We have made a very big step: we have replaced the QCD 
lagrangian ${\cal L}_\mathrm{QCD}$ 
by the effective lagrangian ${\cal L}_\mathrm{eff}$. 
This transition may be visualized in terms of Feynman graphs:
in Fig.~\ref{fig:pipiqcd} 
 is displayed one of the infinitely
many graphs that contribute to $\pi\pi$ 
scattering in QCD, and which are all equally important.
On the other hand, in the effective theory, only the two graphs
displayed in Fig.~\ref{fig:pipieff}
 contribute at leading order. In the figure, the 
symbol $M^2$ stands for the combination
\bea\label{eq:M2}
M^2=(m_u+m_d)B
\eea 
which finally counts in ${\cal L}_2$, see below. 
The crucial point to observe is the fact that the  transition 
\bc
${\cal L}_\mathrm{QCD}\Rightarrow {\cal L}_\mathrm{eff}$
\ec
is a non perturbative phenomenon. It is very different 
from the construction of the effective theory 
for the linear sigma model,
 where the loop expansion in the original 
theory does make sense, and where the low-energy 
representation  can be worked out order by order 
in the loop expansion.

I have provided 
\begin{enumerate}
\item[\bul]
no proof that ${\cal L}_2$ is 
the correct leading order lagrangian.
\item[\bul]
no discussion of the group theoretical background 
(realization of the group $SU_R(2)\times SU_L(2)$ 
on curved manifolds).
\end{enumerate}
For these issues, I refer the 
interested reader to earlier Schladming lectures by 
Leut\-wyler~\cite{leutwylerschladming}, 
Manohar\cite{manoharschladming} and  
Ecker\cite{eckerschladming}. Leutwyler has 
proven in \cite{leutwylerfoundations} that the above 
lagrangian does reproduce the Green functions of QCD at 
low energy, see also below.

\setcounter{equation}{0}
\section{Calculations with ${\cal L}_\mathrm{eff}$}
In the last section, we have displayed the effective 
lagrangian that allows one to evaluate masses, form 
factors and scattering matrix elements in QCD at low energy.
 The results are valid 
at small values of the quark masses and of the momenta.
In this section, I illustrate the procedure with 
several examples.
\subsection{Leading terms}
Evaluating matrix elements at tree level with the 
effective lagrangian  (\ref{eq:qcdleading})
generates the leading term in the low-energy expansion, 
see later for a precise meaning of  this terminology.
\subsubsection*{Pion masses}


Consider the kinetic term in ${\cal L}_2$,
\bea
{\cal L}_2&=&\frac{1}{2}[\partial_\mu
{\vec{\pi}}
\cdot\,
\partial^\mu
{\vec{\pi}} 
-M^2
{\vec{\pi}}^2]
+O(
{\vec{\pi}}^4
)\fs
\neea
We conclude that the three  pions have the same mass in 
this approximation,
\bea\label{eq:pionmass1}
M_{\pi^\pm}^2=M_{\pi^0}^2=M^2\fs
\eea
{\underline{Remark:}} 
 We denote by $M_{\pi^\pm}$ and 
$M_{\pi^0}$ the charged and neutral pion mass, respectively.
 The symbol $M_\pi$ stands for the   pion mass 
 in the isospin symmetry limit, whereas 
 $M^2$ is the first term in
 the quark mass expansion of the charged and neutral 
 pion (mass)$^2$.

\subsubsection*{$\pi\pi$ scattering}
The terms of order $\vec{\pi}^4$ in ${\cal L}_2$ generate the 
leading term in the $\pi\pi$ scattering amplitude,
\bea
{\cal L}_2&=&\mbox{kinetic term}\, +\, 
\frac{1}{8F^2}[\partial_\mu{\vec{\pi}}^2
\partial^\mu{\vec{\pi}}^2
-M^2{\vec{\pi}}^4
]+O({\vec{\pi}}^6)\fs
\neea
The couplings are 
\bea
\frac{1}{F^2}\,\,\mbox{and}\,\,\frac{M^2}{F^2}\fs
\neea
In other words, the mass $M$ is also a coupling! 
Figure~\ref{fig:pipieff} displays the two diagrams 
that one has to evaluate in this case. 

\vskip.5cm
\bc
{\textbf { a) $\pi^+\pi^-\rightarrow \pi^0\pi^0$}}
\ec
\vskip.5cm

 The scattering matrix element is
\bea
T&=&\frac{M^2-s}{F^2}+O(p^4)\sem s=(p_1+p_2)^2\fs
\neea
This result agrees with (\ref{eq:tcharged}), 
up to the replacement
 $(M,F)$$\rightarrow (M_\pi,F_\pi)$, 
which is a higher order effect, included in the symbol 
 $O(p^4)$.

\vskip.5cm

\bc
{\textbf { b) $\pi^a\pi^b\rightarrow \pi^c\pi^d$}}
\ec

\vskip.5cm

The structure of the matrix element  
in the general case is displayed
in (\ref{eq:Apipi}). The invariant amplitude $A(s,t,u)$ becomes
\bea\label{eq:A_2chpt}
A(s,t,u)&=&\frac{s-M^2}{F^2}+O(p^4)\fs
\eea

\vskip.5cm

\bc
{\textbf { c) Isospin amplitudes}}
\ec

\vskip.2cm

For later use, we introduce here the isospin amplitudes 
\bea
T^{I=0}&=&3A(s,t,u)+A(t,u,s)+A(u,s,t)\co\nn
T^{I=1}&=&A(t,u,s)-A(u,s,t)\co\nn
T^{I=2}&=&A(t,u,s)+A(u,s,t)\fs
\neea

\subsection{Higher order trees}
Evaluating tree-level graphs with ${\cal L}_2$ generates the leading 
terms of the quantities in question. How about the 
contributions from ${\cal L}_4$? It contains e.g. terms with 
four derivatives, like
\bea
{\cal L}_4=d_1\la\partial_\mu U\partial^\mu 
U^\dagger\ra^2+\cdots\fs
\neea
What is the effect of this term?
From
\bea
\la\partial_\mu U\partial^\mu U^\dagger\ra^2=\frac{4}{F^4}
(\partial_\mu{\vec{\pi}}
\cdot\,\partial^\mu{\vec{\pi}}
)^2+\cdots\co
\neea
we conclude that ${\cal L}_4$  also contributes to 
elastic $\pi\pi$ 
scattering. Since four derivatives are involved, 
there will be four powers of the momenta. Indeed,
at tree level, ${\cal L}_4$ contributes with
\bea\label{eq:pipi4}
\delta A(s,t,u)=
\frac{8d_1}{F^4}(s-2M^2)^2+\cdots\fs
\eea
This term is of order $p^4$ for small momenta.
 One can  perform the expansion in a systematic manner 
- there are only a limited number of 
terms at order $p^4$ \cite{annals}:
\bea
{\cal L}_4&=&\sum_{i=1}^7l_iP_i\co\nn
P_1&=&\frac{1}{4}\la\partial_\mu U\partial^\mu 
U^\dagger\ra^2\scs
P_2=\frac{1}{4}\la\partial_\mu U \partial_\nu U^\dagger\ra
    \la\partial^\mu U \partial^\nu U^\dagger\ra\co\nn
P_3&=&\frac{1}{16}\la 2B{\cal M}(U+U^\dagger)\ra^2\scs
P_7=-\frac{1}{16}\la 2B{\cal M}(U-U^\dagger)\ra^2\fs\nn
\neea
$P_{4,5,6}$ do not contribute to the $\pi\pi$ scattering 
amplitude.
Why is knowledge of ${\cal L}_4$ useful? Suppose that the LECs
$l_i$ are known 
$\Rightarrow$ low-energy amplitudes 
are parametrized in terms of 7 parameters $\Rightarrow$ 
all other amplitudes fixed in terms of these 
parameters at this order in the low-energy expansion.
\subsubsection*{Comments}
\begin{itemize}
\item[\bul]
Chiral symmetry and $C,P,T$ invariance  have
 been used to determine ${\cal L}_4$.
For example,
\bea
\la \partial_\mu U\partial^\mu U^\dagger\ra^2
\neea
is invariant under $
U\rightarrow V_RUV_L^\dagger\scs V_I\in SU(2)\fs$
\item[\bul]
The low-energy constants $l_i$ (LECs) are not fixed by 
 symmetry arguments.  
\item[\bul]
The operators $P_{1,2,3,7}$ contribute 
to the following processes:
\bea
\begin{tabular}{lcl}
$P_1,P_2$&$\rightarrow$&$\pi\pi\rightarrow\pi\pi$,\,\ldots\\
$P_3$ &$\rightarrow$&$M_\pi^2,\pi\pi\rightarrow\pi\pi$,\,
\ldots\\
$P_7$&$\rightarrow$&$M_{\pi^+}^2-M_{\pi^0}^2\fs$
\end{tabular}
\neea
\end{itemize}

\subsubsection*{Example }
Taking into account the contributions from ${\cal L}_4$,
 the pion masses read at tree level
\bea\label{eq:piontreep4}
M_{\pi^\pm}^2&=&M^2+\frac{2l_3}{F^2}M^4\co\nn
M_{\pi^0}^2&=&M_{\pi^\pm}^2-\frac{2B^2}{F^2}(m_d-m_u)^2l_7
\fs
\eea

\subsection{Loops}
Scattering amplitudes, evaluated with
\bea
{\cal L}_2+{\cal L}_4
\neea
at tree level, are still real.
 To be in accord  with optical theorem, one needs to 
evaluate loops. This guarantees that unitarity is satisfied
order by order in the low-energy expansion, like in any 
 standard loop expansion in QFT. We illustrate the 
evaluation of loops in the case of the pion mass.

\subsubsection*{Pion mass}
To evaluate the pion mass in the isospin symmetry 
 limit $m_u=m_d$,
 we consider the connected 
two-point function
\bea\label{eq:twopoint}
\delta^{ab}\triangle_c(p)=i\int d^4x\,e^{ipx}
\la0|T\phi^a(x)\phi^b(0)|0\ra_c\fs
\eea
\begin{figure}[H]
\bc
\epsfig{figure=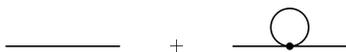,
width=6cm,
bbllx=207,
bblly=690,
bburx=380,
bbury=735,
clip=
}
\caption{Tree and tadpole contribution from ${\cal L}_2$ 
to the two-point function ({\protect{\ref{eq:twopoint}}})
}\label{fig:pionmass}
\ec
\end{figure}
Taking into account the tree and tadpole contributions 
displayed in 
Fig.~\ref{fig:pionmass},  we find that
\bea
\triangle_c(p)=\frac{Z}{M_p^2-p^2}\co
\neea
where
\bea
M_p^2&=&M^2+\frac{M^2I}{2F^2}\scs
I=\int\frac{d^4l}{i(2\pi)^4}\frac{1}{M^2-l^2}\fs
\neea
The tadpole integral $I$ is quadratically divergent. 
A standard method to cope
with this situation is to perform the integral 
in $d$ dimensions,
\bea
I\rightarrow&& \int\frac{d^dl}{i(2\pi)^d}\frac{1}
{M^2-l^2}
=\frac{M^{d-2}}{(4\pi)^{d/2}}\Gamma(1-d/2)\fs
\neea
The result is now finite at $d\,\neq 2,4,6\ldots$, whereas 
 it is still still divergent at $d=4:$
\bea
{M^2I}\rightarrow \frac{M^4}{8\pi^2}
\frac{1}{d-4}\scs d\rightarrow 4\fs
\neea
There  is also a tree contribution from ${\cal L}_4$, see
 (\ref{eq:piontreep4}),
\bea
\delta M_p^2=\frac{2l_3}{F^2}M^4\fs
\neea
If we tune $l_3$ to diverge as $d\rightarrow 4$ as well, 
we may cancel
 the divergence in the mass:
\bea
l_3\rightarrow -\frac{1}{32\pi^2}\frac{1}{d-4}
+\mbox{finite}\scs 
d\rightarrow 4
\neea
The physical pion mass at $m_u=m_d$ finally becomes
\bea\label{eq:pionmassqcd}
M_\pi^2=
M^2-\frac{M^4}{32\pi^2F^2}\bar l_3 + O(M^6)\fs
\eea
The quantity ${\bar l}_3$ contains the finite parts from 
$l_3$ and  from the tadpole integral $I$ \cite{annals}.
\subsubsection*{Comments}
\begin{itemize}
\item
The required counterterm to cancel the divergence
stems from ${\cal L}_4$. The divergences cannot be canceled 
by tuning parameters  in ${\cal L}_2$: this is a 
typical feature of a  non-renormalizable theory.
\item
Non-renormalizability does not mean {  non-calculability:}
 the effective theory of
QCD allows one to calculate all quantities 
perfectly well. 
 The only disadvantage is the
 increasing number of low-energy coupling constants 
that one is faced with
 while incorporating  higher order terms in 
the expansion.
\item
The contribution from the tadpole is suppressed, 
it is of order $M^4$, not of order $M^2$.
\end{itemize}
\subsection{Renormalization made systematic}
Similar divergences occur in other amplitudes, like 
$\pi\pi\rightarrow\pi\pi$,
when loops are calculated.
These divergences can {all} be canceled by tuning 
the $l_i$ in the lagrangian with four derivatives. The final 
prescription for the evaluation of the Green functions at 
order $p^4$ reads as follows: the effective lagrangian is
\bea
{\cal L}_\mathrm{eff}={\cal L}_2+{\cal L}_4+{\cal L}_6
+\cdots\co
\neea
where ${\cal L}_2$ is given in (\ref{eq:qcdleading}), and
\bea
{\cal L}_4&=&\sum_{i=1}^7 l_iP_i\scs
{\cal L}_6=\sum_{i=1}^{53} c_i Q_i\fs
\neea
The polynomials $P_i\, (Q_i)$ and the divergent parts 
 in the LECs $l_i\, (c_i)$ have been determined in
 \cite{annals}\,\,(\cite{p6bce}). At order $p^6$, 
there are in addition so 
called  parity odd terms, that come with an epsilon 
tensor~\cite{p6odd}. 
Further, some of the polynomials $P_i,Q_i$ vanish 
in the absence of external fields. I refer the reader 
to \cite{p6bce} 
 for a guided tour through ${\cal L}_6$.

The Green functions are calculated as follows:

\vskip1cm

leading terms:\hspace{5mm}trees with ${\cal L}_2$
\bea
\mbox{next-to-leading terms: \hspace{2mm}}
\biggl\{\begin{tabular}{ll}
trees with &${\cal L}_4$\\
one loop with &${\cal L}_2$
\end{tabular}
\neea
\bea
\mbox{next-to-next-to-leading terms:\hspace{2mm}}
\biggl\{\begin{tabular}{ll}
trees with &${\cal L}_6$\\
one loop with &${\cal L}_4$\\
two loops with &${\cal L}_2$\\
\end{tabular}
\neea

\vskip1cm
\subsubsection*{Comments}
\begin{enumerate}
\item[\bul]
Next-to-leading terms are suppressed by $p^2$ with respect 
to the leading term, the next-to-next-to leading terms by 
$p^4$, etc.
\item[\bul]
The finite parts of the low-energy constants $l_i,c_i$ are 
not fixed by symmetry
considerations. On the other hand, they  are calculable 
in  principle in QCD.
We are witnessing a transition period, where one 
starts to be able to evaluate the LECs in the framework of QCD
 - see \cite{lattlecs} for  recent  examples.
\item[\bul]
Meanwhile, until the transition period is over, one 
 takes LECs from data, from sum rules etc.
\item[\bul]
We will comment in the final section about 
the introduction of additional
fundamental fields in the effective lagrangian. 
\end{enumerate}

\setcounter{equation}{0}
\section{Pion-pion scattering}
In this section, I consider in some detail the evaluation of 
the elastic $\pi\pi$ scattering amplitude in the 
low-energy region at order $p^6$. The motivation 
for investigating this reaction in great detail 
includes the following points:
\begin{enumerate}
\item[\bul]
It is possible to make precise theoretical predictions.
\item[\bul]
Some of the threshold amplitudes are sensitive to 
the mechanism of spontaneous chiral symmetry breaking.
\item[\bul]
There are new experimental activities:

\begin{tabular}{lll}
BNL-865 at Brookhaven $\,\,\,$ &$K_{e4}$ decays&
\protect{\cite{E865}}\\
NA48 at CERN &$K_{e4}$ decays&\protect{\cite{na48}}\\
KLOE at DAFNE &$K_{e4}$ decays&\protect{\cite{handbook}}\\
DIRAC at CERN &$\pi^+\pi^-$ atom&\protect{\cite{dirac}}
\end{tabular}
\item[\bul]
The analysis of the $P$-wave amplitude in 
elastic $\pi\pi$ scattering provides 
very useful information for the evaluation of the 
anomalous magnetic moment 
of the muon~\cite{leutwylerff}.
\end{enumerate}

Why do the so called $K_{e4}$ decays 
\bea
K^+&&\rightarrow\pi^+\pi^-e^+\nu_e\co\nn
K^+&&\rightarrow\pi^0\pi^0e^+\nu_e\co\nn
K^0&&\rightarrow\pi^0\pi^-e^+\nu_e\co
\neea
and their charge conjugate modes 
provide information on $\pi\pi$ scattering?

\subsubsection*{First explanation} 
The emerging pions in the decay products 
interact with each other ({\em final state interaction}). 
The decay width is therefore sensitive to this interaction.

\subsubsection*{Second explanation (more learned)}
The decay matrix element is described by form factors. One may 
perform a 
partial wave analysis of these - the partial wave 
amplitudes then carry the phases of the
 $\pi\pi$ interaction (Watson's final state theorem). In the 
decay, one
measures interference terms between the various partial waves.
The decay of the charged kaon into a charged pion pair
plus leptons  turns out to be  
sensitive to
\bea
\delta_{l=0}^{I=0}(E_{\pi\pi})-\delta_1^1(E_{\pi\pi})\co
\neea
where $E_{\pi\pi}$ denotes the centre-of-mass energy 
of the pion pair.
The investigation of $\pi^+\pi^-$ atoms (Pionium) 
is of interest for the
following reason. The atom is formed by electromagnetic 
interactions. It is not stable - the ground state, e.g.,  
 has various decay channels,
\bea
A_{\pi^+\pi^-}\rightarrow \pi^0\pi^0,\gamma\gamma,
\pi^0\pi^0\gamma\gamma,\ldots\fs
\neea
The decay into the neutral pion pair depends on the 
strength of the strong interactions~\cite{deser,boundstates},
\bea\label{eq:deser}
\Gamma=\frac{2}{9}\alpha^3p^*|a_0^0-a_0^2|^2(1+\delta)\co
\eea
where $a_0^0$ denotes the $I=0$ $S$-wave scattering length 
$a_{l=0}^{I=0}$, 
 and $p^*$ is the modulus of the 
centre-of-mass momentum of the 
neutral pions in the rest system of the decaying atom. Further,
$\alpha\simeq 1/137$ is the fine structure constant of QED.
 The decay width into a photon 
pair is suppressed by the factor 
 $\alpha^{3/2}$ with respect to the $\pi^0\pi^0$ channel.
The correction $\delta$ is also known~\cite{boundstates},
\bea
\delta=0.058\pm0.012\fs
\neea
A measurement of the lifetime of the ground 
state therefore amounts to a 
measurement of the combination $|a_0^0-a_0^2|$.

\subsection{Chiral expansion of the $\pi\pi$ amplitude}
 The chiral expansion of the invariant scattering 
amplitude $A(s,t,u)$ is performed in the manner 
discussed in the  previous section and has the structure
\bea
A(s,t,u)&=&A_2+A_4+A_6+\cdots\co
\neea
where $A_{2n}$ is of order $p^{2n}$. 
\subsubsection*{Leading order}
 The leading order result is given in (\ref{eq:A_2chpt}).
 For the $I=0$ $S$-wave scattering length, one obtains from 
that expression
\bea\label{eq:pipifirst}
a_0^0=\frac{7 M^2}{32\pi F^2}=0.16\fs
\eea
In the numerical evaluation, I have replaced  $M$ and $F$ 
 by $M_{\pi^\pm}$ = 139.57 MeV  
and by 
$F_\pi=92.4$ MeV, respectively (the replacements 
amount to taking into
 account some of the higher
 order effects in the calculation of the 
scattering amplitude). The data on $K_{e4}$ decays 
 collected in the seventies gave~\cite{rosselet} 
\bea
a_0^0=0.26\pm 0.05 \qquad 
\mbox{[from 30'000 $K_{e4}$ decays]}\fs
\neea
The question was -- for many years -- 
 whether this indicates a 
failure of the chiral prediction (\ref{eq:pipifirst}).
Indeed, if the condensate $\la0|\bar u u|0\ra$ 
is small or vanishing, one
can understand a large value of the scattering length.
 This is the main idea of the 
so-called {\em generalized chiral perturbation
  theory}, which was much discussed at the end of 
the last millennium \cite{gchpt}.
In order to decide the issue, one needs more 
precise data and a more precise calculation. 
 Both is available in our days, as we will show below.

\subsubsection*{Next-to-leading order}
\begin{figure}[H]
\bc
\epsfig{figure=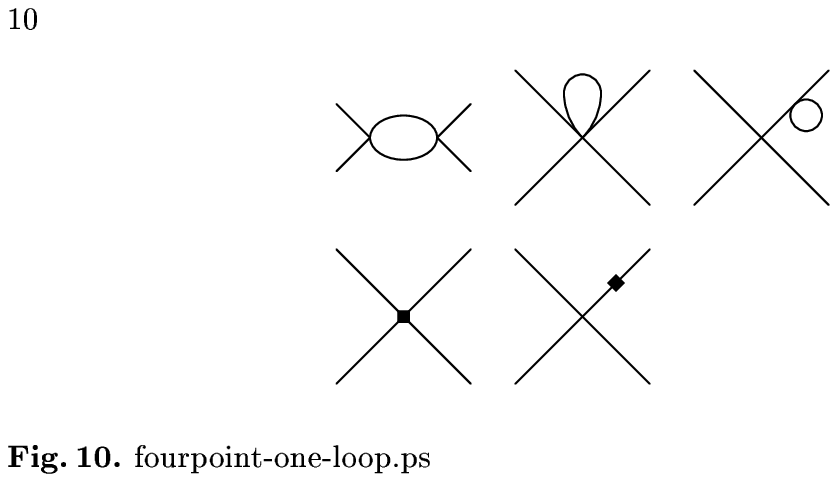,
width=8cm,
height=4cm,
bbllx=207,
bblly=627,
bburx=406,
bbury=741,
clip=
}
\caption{One-loop graphs in elastic $\pi\pi$ scattering. 
The {\it filled
  squares} denote contributions from ${\cal L}_4$}
\label{fig:onelooppipi}
\ec
\end{figure}

One evaluates 
one-loop graphs with ${\cal L}_2$ and tree-graphs 
with ${\cal L}_4$. Some of these are displayed in 
Fig.~\ref{fig:onelooppipi}.
The result can be written in the form
\bea
A_4&=&a_1M_\pi^4+a_2M_\pi^2s+a_3s^2+a_4(t-u)^2\nn
&&+F(s)+G(s,t)+G(s,u)\fs
\neea
We note the following:
\begin{enumerate}
\item[\bul]
The amplitude $A_4$ carries four  powers of momenta
\item[\bul]
$F,G$ denote loop functions. They generate the imaginary parts 
which are required by unitarity at order $p^4$. Example:
\bea
F(s)&=&\frac{1}{2F_\pi^4}(s^2-M_\pi^4)\bar J(s)\co\nn
\bar
J(s)&=&\frac{1}{16\pi^2}\biggl\{\sigma
\log{\frac{1-\sigma}{1+\sigma}}+2+i\pi\sigma\biggr\}\co\nn
\sigma&=&(1-{4M_\pi^2}/{s})^{1/2}\sem s > 4 M_\pi^2\fs
\neea
 The 
function $\bar J (s)$ is analytic in the complex $s$-plane, 
cut along the real axis for $s\geq 4M_\pi^2$.
\end{enumerate}
\begin{enumerate}
\item[\bul]
$a_i$ contain the low-energy constants $\bar l_{1,2,3,4}$ 
from ${\cal L}_4$.
 For example, the $I=0$ $S$-wave scattering length now reads
\bea
a_0^0&=&\frac{7}{32\pi}\frac{M_\pi^2}{F_\pi^2}\biggl\{
1+\varepsilon+O(M_\pi^4)\biggr\}\co\nn
\varepsilon&=&\frac{5}{84\pi^2}\frac{M_\pi^2}{F_\pi^2}\biggl
(\bar l_1+2\bar l_2-\frac{3}{8}\bar
 l_3+\frac{21}{10}\bar l_4+\frac{21}{8}\biggr)\fs
\neea
\item[\bul]
the LECs $\bar l_i$ contain so-called 
{  \em{chiral logarithms}}:
\bea
\bar l_i\rightarrow 
-\log{M_\pi^2}\scs M_\pi\rightarrow 0\co
\neea
which generate large corrections,
\end{enumerate}
\bea
a_0^0&=&\frac{7M_\pi^2}{32\pi F_\pi^2}
\biggl\{1-
\underbrace{
\frac{9}{32\pi^2}\frac{M_\pi^2}{F_\pi^2} \log{M_\pi^2/\mu^2}
}_{\chi\mbox{logarithm}}
+\cdots\biggr\}\fs
\neea
The $\chi$ logarithms generate
 a $25\%$ correction at $\mu=1$ GeV.
\subsubsection*{Next-to-next-to leading order}
\begin{figure}[H]
\bc
\epsfig{figure=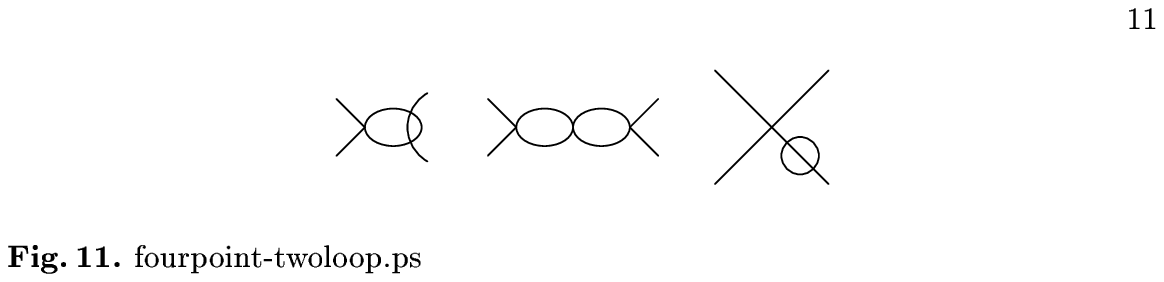,
width=8cm,
height=2.5cm,
bbllx=215,
bblly=687,
bburx=401,
bbury=744,
clip=
}
\caption{Two-loop graphs in elastic $\pi\pi$
  scattering}\label{fig:twoloopspipi}
\ec
\end{figure}

In order to evaluate $A_6$, one needs to perform a two-loop
calculation with ${\cal L}_2$, one-loop with ${\cal L}_4$ 
and trees with ${\cal L}_6$. A few of the two-loop graphs 
are displayed in Fig.~\ref{fig:twoloopspipi}.
It turns out that all graphs can be evaluated 
in closed form~\cite{Atwoloops}!
 The structure of $A_6$  is as follows:
\bea
A_6&=&\mbox{loop functions}\nn
&&+b_1M_\pi^6+b_2M_\pi^4s+b_3M_\pi^2s^2
+b_4M_\pi^2(t-u)^2 + b_5s^3+b_6s(t-u)^2\fs
\neea
The calculation determines $b_1,\ldots, b_6$ in 
terms of the LECs at order $p^4$ and $p^6$~\cite{Atwoloops}.
Obviously, one is faced with a problem here: one 
needs to determine the LECs
before a precise prediction of the scattering lengths 
can be performed.

\subsection{Type A,B LECs}\label{sec:subsectypealecs}
One encounters the following LECs in the $\pi\pi$ 
amplitude up to and including terms of order $p^6$:

\begin{tabular}{ll}
\hspace{2cm}&\\
 $\bar{l}_1,\bar{l}_2,\bar{l}_3,\bar{l}_4$
&:\hspace{.2cm} ${\cal L}_4$\\
$ \bar{r}_1,....,\bar{r}_6$&:\hspace{.2cm} ${\cal L}_6$\\
\end{tabular}

They come in two categories~\cite{pipicgl}:

{\em 1. Terms that survive in the chiral limit}:
\begin{center}

  {$\bar{l}_1,\bar{l}_2,\bar{r}_5,\bar{r}_6$} \, {  type A}

\end{center}
These LECs show up in momentum dependence 
of the $\pi\pi$ amplitude,
 and  may therefore be determined phenomenologically.

{\em 2. Symmetry breaking terms}: 

\begin{center} {$\bar{l}_3,\bar{l}_4,\bar{r}_1,\bar{r}_2,
\bar{r}_3, \bar{r}_4$}\,\, {  type B}
\end{center}
These LECs specify the quark mass dependence of the amplitude.
 The $\pi\pi$ scattering amplitude cannot provide  
information on type B LECs, because one cannot vary the 
quark mass in experiments. 

We discuss in the following section how these LECs can 
be determined, and what the result for the scattering 
lengths finally is.

\setcounter{equation}{0}
\section{Roy equations and threshold parameters}
Roy equations allow one to determine type A LECs and to 
pin down the
$\pi\pi$ scattering amplitude in the low-energy 
region with high precision. It is useful to discuss Roy 
equations in
two steps. In the first step, no relation to chiral 
symmetry is used - the only ingredients are analyticity,
 unitarity,
crossing symmetry,  data above  800 MeV and Regge 
behaviour beyond 2 GeV.
 In a second
step, one merges the representation of the amplitude with 
the chiral representation discussed in the previous section.

\subsection{$\pi\pi\rightarrow\pi\pi$: Roy  I}

\begin{figure}[H]
\bc
\includegraphics[angle=-90,width=8cm]{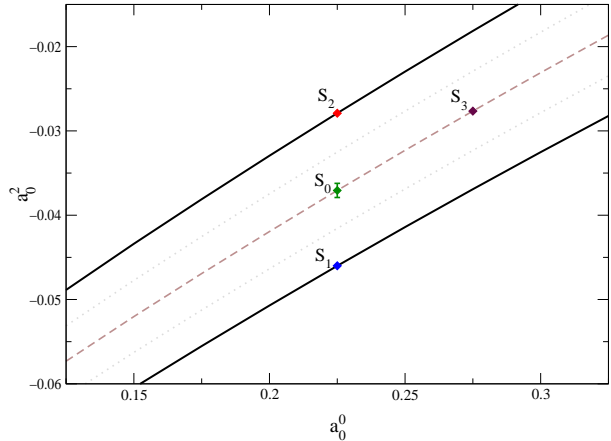}
\caption{The universal band in the $a_0^0-a_0^2$ plane. For 
each point
  in this band, a unique solution to the Roy equations
 can be constructed}
\label{fig:universal}
\ec
\end{figure}

\begin{figure}[H]
\bc
\includegraphics[angle=-90,width=7cm]
{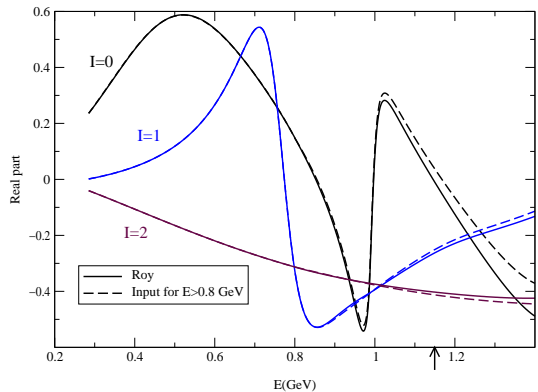}
\caption{
Example of a solution
to the Roy equations. Displayed are the real parts of 
the partial waves, as a function of the centre-of-mass 
energy of the pions}\label{fig:solution}
\ec
\end{figure}

Analyticity, crossing symmetry and the Froissart bound lead to 
dispersion relations for the partial waves of the 
$\pi\pi$ scattering amplitude. As has been shown by 
Roy~\cite{roy}, the imaginary part of the partial waves 
is needed only in the physical region. These dispersion 
relations read
\bea t_\ell^I(s)= k_\ell^I(s)+
\sum_{I'=0}^2\sum_{\ell'=0}^\infty \int_{4M_\pi^2}^\infty
ds'\,K_{\ell\ell'}^{I I'}(s,s')\,\mbox{Im} \,
 t_{\ell'}^{I'}(s')\co \nonumber
\eea
where $\ell (I)$ denotes the angular momentum (isospin) of 
the partial wave. The subtraction term
$k^I_\ell(s)$ contains the scattering lengths $a_0^0,a_0^2$, 
and the kernels $K_{\ell\ell'}^{I I'}(s,s')$ 
are explicitly known functions of $s$ and of $s'$.
 At low energy, only $S$- and $P$- waves matter, and the
 contributions from the remaining waves may be  expanded in a 
Taylor series.
 Unitarity expresses the imaginary parts of the partial waves 
in terms of the 
real parts (in the elastic region) -- the above dispersion 
relations then become (singular) integral equations for the 
two $S$- and for the $P$-wave.
 These may be solved numerically~\cite{solveroy,physrep}.
In~\cite{physrep},
 it has been shown  that
 the two $S$-wave scattering lengths $a_0^0,a_0^2$ are 
the essential parameters. Once these are fixed,
experimental data (above 800 MeV) plus Regge behaviour 
above 2 GeV determine the amplitude in the low
energy region to within rather small uncertainties.
Figure \ref{fig:universal} displays the {\em universal band} 
 in the $a_0^0-a_0^2$ plane, for which solutions may be found.
 An example solution for the three waves is displayed  
in Fig.~\ref{fig:solution}.

\vskip.5cm


\subsection{$\pi\pi\rightarrow\pi\pi$: Roy II}
\begin{figure}[H]
\bc
\includegraphics[width=8cm]{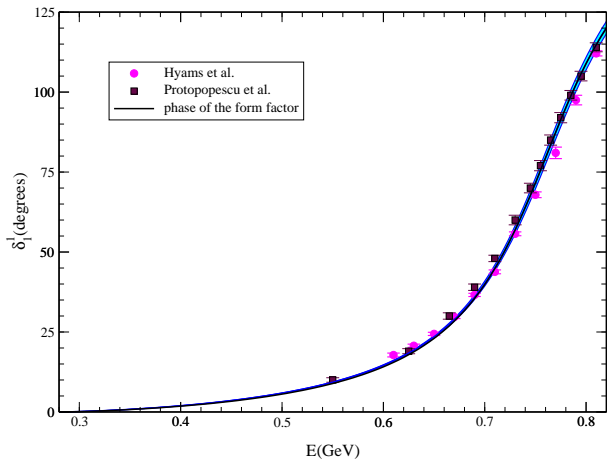}
\caption{The $P$-wave from the Roy analysis}\label{fig:pwave}
\ec
\end{figure}

In the previous subsection, no reference to ChPT was made.
 We now invoke this information: one requires that 
the $\chi$ amplitude 
agrees with phenomenological amplitude
below the threshold region. Subtractions and matching 
 are performed in such a manner  that $\chi$ logarithms 
are  suppressed. This procedure allows one to determine the 
type A LECs listed in subsection \ref{sec:subsectypealecs}.
Type B LECs are determined form other sources:
\bea
\begin{tabular}{ll}
$\bar{l}_4: \leftrightarrow$& scalar radius of pions\\
$\bar{l}_3: \leftrightarrow$& SU(3), Zweig rule \\
$\bar{r}_{1,2,3,4}\leftrightarrow$ &resonance saturation
\end{tabular}
\neea
We have now arrived at solutions to the Roy equations 
that agree with the chiral amplitude at low energy. 
 In other words, the three lowest partial waves 
below 800 MeV are fixed. The scattering lengths of the 
partial waves
with $\ell\geq1$, as well as the effective ranges 
(also those of the
$S$-waves) can be expressed in terms of sum rules over 
the imaginary parts~\cite{wanders}. In Fig.~\ref{fig:pwave}
 is displayed the resulting $P$-wave phase shift.
 This allows one to pin down the electromagnetic form
factor of the pion with high precision~\cite{leutwylerff}.
This form factor plays a very important role in the 
evaluation of the anomalous magnetic moment of the 
muon, see Marc Knecht's lectures at this school.

Finally, we come to the two $S$-wave scattering lengths, 
 which are now also fixed~\cite{pipiswave,pipicgl},
\bea
a_0^0&=&0.220 \pm 0.005\co\nn
a_0^0-a_0^2&=&0.265\pm0.004\fs
\neea
I refer the reader to~\cite{pipicgl} for scattering lengths and
effective ranges of other waves. The above result also 
provides~\cite{boundstates}
 a precise prediction for the lifetime of Pionium 
in the ground state via (\ref{eq:deser}),
\bea
\tau&=&(2.9\pm0.1)\times10^{-15}\,{\mbox{sec}}\fs
\neea
This prediction is presently confronted with experiment
 at DIRAC~\cite{dirac}.

\subsection{The coupling $\bar l_3$}
\begin{figure}[H]
\bc
\epsfig{file=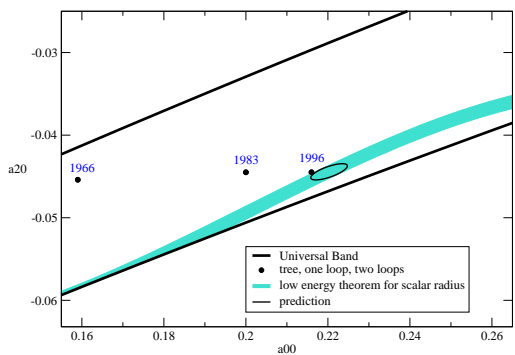,width=6cm,angle=-90}
\caption{Constraints imposed on the $S$-wave scattering lengths
 by chiral symmetry. The three circles illustrate the 
prediction of the chiral perturbation theory at 
increasing order. The
error ellipse represents the final result from 
Ref.~\cite{pipicgl}, while the narrow, curved band 
indicates the
region allowed in GChPT}\label{fig:bandl3}
\end{center}
\end{figure}

The main difference between generalized chiral 
perturbation theory (GChPT \cite{gchpt}) 
and the standard picture used here resides in the 
coupling constant ${\bar l}_3$, which can take any 
value in GChPT. Let me recall where this constant occurs. 
First, consider the
chiral expansion of the pion mass (\ref{eq:pionmassqcd}).
 We write
\bea
\bar{l}_3 = \log{\frac{\Lambda_3^2}{M_\pi^2}}\fs
\neea
Crude estimates in the standard version of ChPT 
give~\cite{annals}
\bea
       0.2\,\,  {\mbox{GeV}} < \Lambda_3 <  2\,\,
 {\mbox{GeV}}\fs
\neea
The term  of order $M^4$ in (\ref{eq:pionmassqcd}) 
is then very small compared to the leading term,  
 i.e., the Gell-Mann-Oakes-Renner formula is obeyed 
very well.
 As mentioned, GChPT allows for arbitrarily large 
values of $\bar{l}_3$. The quadratic term in 
(\ref{eq:pionmassqcd}) is then 
not leading, the series must be reordered. 
 It is very satisfactory that experiment can decide the issue,
 for the following reason.
 The constant  ${\bar l}_3$ also occurs in the
expression for the scattering lengths $a_0^0$ and $a_0^2$.
 One may then perform the matching of the chiral and the
 phenomenological amplitude as discussed above 
 with $\bar{l}_3$ as a free parameter. The result of this
 investigation~\cite{pipicgl,kl4cgl} is displayed 
in Fig.~\ref{fig:bandl3}~\cite{leutwylerfigure}. 
The allowed values of the scattering lengths lie in the 
small hatched band in the figure. This band reflects
 a low-energy theorem~\cite{pipicgl}
 for the difference $2a_0^0-5 a_0^2$,
\bea\label{eq:lowenergytheorem}
2a_0^0-5a_0^2=\frac{3M_\pi^2}{4\pi F_\pi^2}\biggl(
1+\frac{M_\pi^2\la
    r^2\rangle_s}{3}+\frac{41 M_\pi^2}{192\pi^2F_\pi^2}
+O(M_\pi^4)\biggr)\co
\eea
where $\la r^2\rangle_s$ denotes the scalar radius of the pion.
 The terms of order $M_\pi^6$ in (\ref{eq:lowenergytheorem}) 
generate the curvature of the band.
For each value in the band, one can calculate the 
difference $\delta_0^0-\delta_1^1$
of $\pi\pi$ phase shifts, and compare with data on
 $K^+\rightarrow \pi^+\pi^-e^+\nu_e$ decays.
E865 has performed this analysis, with the result~\cite{E865}
\bea
a_0^0=0.216\pm 0.013\, ({\mbox {stat}}) \pm 0.002
\, ({\mbox {syst}}) 
\pm 0.002 \, ({\mbox {theor}})\fs 
\neea
The central value leads to $\bar{l}_3 \simeq 6$,
 with an uncertainty of about 10 units. In  
the expansion (\ref{eq:pionmassqcd}),
the second term  amounts to a correction 
of about four percent at $\bar l_3$ =6.
From this, one concludes that the first term in the 
mass expansion 
dominates by far:
 the motivation for a generalized scheme~\cite{gchpt}, 
with a small quark condensate and a large second term in
(\ref{eq:pionmassqcd}) has evaporated, 
at least for $SU(2)\times SU(2)$.

\subsection{Note added after the lectures}

\subsubsection*{On the precision of the theoretical
  predictions for $\pi\pi$ scattering}

In two recent papers \cite{py}, Pel\'aez and Yndur\'ain evaluate 
some of the low energy
observables of $\pi\pi$ scattering and obtain flat 
disagreement with our
results \cite{pipicgl} that I have described above.
 The authors work with unsubtracted 
dispersion relations,
so that their results are very sensitive to the poorly 
known high energy
behaviour of the scattering amplitude. They claim that 
the asymptotic
representation we used in \cite{physrep,pipicgl} is incorrect 
and propose an alternative one. 
We have repeated \cite{pipica} their calculations on the 
basis of the standard, 
subtracted
fixed-$t$ dispersion relations, using their asymptotics. 
The outcome fully
confirms our earlier findings. Moreover, we show that the
Regge parametrization proposed by these authors for the 
region above 1.4 GeV
violates crossing symmetry: Their ansatz is not consistent 
with the
behaviour observed at low energies.

\setcounter{equation}{0}
\section{Outlook}

Instead of a summary, I have provided at the school an 
outlook on topics not covered in the lectures. This outlook 
was based on the structure of the effective chiral 
lagrangian of the Standard Model~\cite{eckersm} displayed 
 in Table 1. The numbers in brackets denote the number of 
independent LECs$\,$\footnote{In ${\cal L}_{p^4}^{\pi N}$, 
I changed 
the number $114$ in Eckers table into $118$, 
 in order to agree with {\protect{\cite{piN4}}}. 
I thank Ulf-G.~Mei\ss ner for pointing out  the correct 
number of LECs in this case.}.
The numbers refer to $N_f=3$, except for the pieces with 
superscript $\pi N$.

{\tiny{
\renewcommand{\arraystretch}{.9}
\begin{table}[ht]
\begin{center}
\vspace{.5cm}
\caption{The effective lagrangian of the Standard 
Model~\cite{eckersm}}
\begin{tabular}{|l|} 
\hline
  \\
\hspace{1cm} ${\cal L}_{\rm chiral\; order}$ 
~($\#$ of LECs) \\[8pt] 
\hline 
 \\
${\cal L}_{p^2}(2)$+${\cal L}_{p^4}^{\rm odd}(0)$+
${\cal L}_{G_Fp^2}^{\Delta S=1}(2)$ +${\cal L}_{e^2p^0}^{\rm em}(1)$
+${\cal L}_{G_8e^2p^0}^{\rm emweak}(1)$  \\[8pt]
~+~${\cal L}_{p}^{\pi N}(1)$+${\cal L}_{p^2}^{\pi N}(7)$
+${\cal L}_{G_8p^0}^{MB,\Delta S=1}(2)$+
${\cal L}_{G_8p}^{MB,\Delta S=1}(8)$ 
+ ${\cal L}_{e^2p^0}^{\pi N,{\rm em}}(3)$  \\[15pt]
~+~$\underline{{\cal L}_{p^4}^{\rm even}(10)}$+
$\underline{{\cal L}_{p^6}^{\rm odd}(32)}$
+$\underline{{\cal L}_{G_8p^4}^{\Delta S=1}(22)}$
+$\underline{{\cal L}_{e^2p^2}^{\rm em}(14)}$ +
$\underline{{\cal L}_{G_8e^2p^2}^{\rm emweak}(14)}$ \\[8pt]
~+~$\underline{{\cal L}_{e^2p}^{\rm leptons}(5)}$ \\[8pt]   
~+~$\underline{{\cal L}_{p^3}^{\pi N}(23)}$+
$\underline{{\cal L}_{p^4}^{\pi N}(118)}$
+ $\underline{{\cal L}_{G_8p^2}^{MB,\Delta S=1}(?)}$ 
+$\underline{{\cal L}_{e^2p}^{\pi N,{\rm em}}(8)}$ 
  \\[15pt]
~+~$\underline{{\cal L}_{p^6}^{\rm even}(90)}$  \\[8pt] 
\hline
\end{tabular}
\end{center}
\end{table}}
}

The underlined lagrangians are fully renormalized: their 
divergence
structure has been determined in a process independent manner.
In the lectures, I had discussed aspects of 
 ${\cal L}_{p^2}, 
{\cal L}_{p^4}^{\rm even}$ and of 
${\cal L}_{p^6}^{\rm even}$ in the $SU(2)\times SU(2)$ case. 
The  above Table opens  a very wide field of 
 further applications.

\subsubsection*{Topics}
Meson and baryon decays:
electromagnetic, semileptonic, leptonic, non leptonic, rare and
not so rare;
 decay constants $F_\pi$, $F_K$, $\ldots $;
scattering amplitudes:
$\pi\pi\rightarrow \pi\pi,\pi N\rightarrow \pi N, 
\gamma N\rightarrow \pi N,\ldots$;
mixing angles;
quark mass ratios;
large $N_C$ investigations;
anomalies;
isospin violation;
weak matrix elements;
quark condensate;
hadronic atoms;
lattice: $m_q\rightarrow 0,V\rightarrow \infty$;
quenched ChPT.

In order to study nuclear physics in the framework of ChPT,
 the above lagrangian must still be enlarged. A vast amount of 
 massive calculations in this
topic have been performed by 
V. Bernard, E.~Epelbaum, W.~Gl\"ockle, N. Kaiser, 
Ulf-G.~Mei\ss ner and others, see e.g. \cite{nuclear}.
This method has allowed  one to put theoretical nuclear 
physics on a sound basis.

For recent contributions to the topics just mentioned, 
see~\cite{jefferson2000,honnef2001}.

\subsubsection*{Acknowledgements} It is a pleasure to thank the
organizers of this Winter School for the warm hospitality, for 
the perfect organization of the School
and for the very pleasant weather conditions.
 In addition, I thank  Christoph H\"afeli and Martin Schmid for
 providing me with many figures that I displayed during the
 lectures, and which are now incorporated partly also here.
Finally, I thank Ulf-G.~Mei\ss ner for carefully reading 
the manuscript.
This work was  supported in part by the Swiss
 National Science Foundation and by RTN, 
BBW-Contract No. 01.0357 
and EC-Contract  HPRN--CT2002--00311 (EURIDICE).

\end{document}